\documentclass[11pt,superscriptaddress,nofootinbib]{article}
\usepackage{cite}
\usepackage{mathptmx}
\usepackage{amsmath,amsfonts,amsthm,amsbsy,amssymb}
\usepackage[small,bf,hang]{caption}
\usepackage[bookmarks=true,hyperfigures=true,hidelinks]{hyperref}
\usepackage{slashed}
\usepackage{latexsym,epsfig}
\usepackage{mathtools}
\usepackage{nicefrac}
\usepackage{mathrsfs}
\usepackage{simplewick}
\usepackage{braket}
\usepackage{bbm}
\numberwithin{equation}{section}
\usepackage{xcolor}
\usepackage{lmodern}
\usepackage{comment}

\def\hybrid{
 \topmargin -20pt
 \oddsidemargin 0pt
 \headheight 0pt \headsep 0pt
 \textwidth 6.25in 
 \textheight 9.5in 
 \marginparwidth .875in
 \parskip 5pt plus 1pt \jot = 1.5ex}

\hybrid
\def\md{\mathrm{d}}
\def\mD{\mathrm{D}}

\def\bec{\begin{center}}
\def\ec{\end{center}}
\def\a{\alpha} 
 
\def\tA{\tilde{A}}

\def\b{\beta}  
\def\c{\gamma} 

\def\d{\delta} 
\def\D{\Delta}

\def\tF{\tilde{\Phi}}

\def\l{\lambda}\def\tl{\tilde{\lambda}}\def\btl{\bar{\tilde{\lambda}}}

\def\m{\mu}
\def\n{\nu}
\def\r{\rho}
\def\s{\sigma}

\def\tG{\tilde{G}}
\def\cG{\mathcal{G}}
\def\pa{{\partial}}

\def\ra{{\rightarrow}}

\def\bE{{\mathbf{1}}}
\def\bX{{\mathbf{X}}}
\def\bY{{\mathbf{Y}}}

\def\tD{\tilde{D}}
\def\cD{\mathcal{D}}
\def\tC{\tilde{C}}
\def\tS{\tilde{S}}
\def\rS{\mathrm{S}}
\def\tX{{\tilde{X}}}

\def\tF{\tilde{F}}

\def\cO{\mathcal{O}}

\def\cG{\mathcal{G}}

\def\cN{\mathcal{N}}

\def\cR{{\mathcal{R}}}
\def\ctR{\mathcal{\tilde{R}}}
\def\tS{\tilde{S}}

\def\cG{\mathcal{G}}

\def\cT{\mathcal{T}}
\def\ctT{\mathcal{\tilde{T}}}

\def\cO{\mathcal{O}}

\def\ra{\rightarrow}

\def\q{\quad} \def\qq{\q\q} \def\qqq{\q\q\q}

\def\tr{{\mathrm{Tr}}}

\begin{document}

\begin{titlepage}
\begin{flushright} 
{\small $\,$}
\end{flushright}
\vskip 1cm
\centerline{\LARGE{\textbf{Perturbative linearization of super-Yang-Mills}}}
\vspace{5mm}
\centerline{\LARGE{\textbf{theories in general gauges}}}
\vskip 1.5cm
\begin{center}
\begingroup\scshape\Large
Hannes Malcha and Hermann Nicolai
\endgroup
\end{center}
\vskip 1cm
\centerline{\textit {Max-Planck-Institut f\"{u}r Gravitationsphysik (Albert-Einstein-Institut)}}
\centerline {\textit {Am M\"{u}hlenberg 1, 14476 Potsdam, Germany}}
\vskip 1.5cm
\centerline{\textbf {Abstract}}
\vskip .3cm
\small{
\noindent
Supersymmetric Yang-Mills theories can be characterized by a non-local and non-linear
transformation of the bosonic fields (Nicolai map) mapping the interacting
functional measure to that of a free theory, such that the Jacobi determinant of the 
transformation equals the product of the fermionic determinants obtained by integrating out 
the gauginos and ghosts at least on the gauge hypersurface. While this transformation 
has been known so far only for the Landau gauge and to third order in the Yang-Mills
coupling, we here extend the construction to a large class of (possibly non-linear
and non-local) gauges, and exhibit the conditions for all statements to remain valid 
off the gauge hypersurface. Finally, we present explicit results to second order in 
the axial gauge and to fourth order in the Landau gauge.
}

\vfill
\end{titlepage}

\section{Introduction}
Pure supersymmetric Yang-Mills theories in $D=3$, $4$, $6$ and $10$ space-time
dimensions \cite{Brink:1976bc} are among the best studied examples
of quantum field theories. Especially the maximally extended $\cN=4$, $D=4$ theory
occupies a central place because of its finiteness properties 
\cite{Mandelstam:1982cb,Brink:1982wv}, its exact quantum conformal invariance,
and its possible relevance for a non-perturbative formulation of string 
theory (M-theory), either via the AdS/CFT correspondence \cite{Maldacena:1997re}
or, in its dimensionally reduced form, via the maximally supersymmetric $D=1$
matrix model with gauge group SU($\infty$) \cite{deWit:1988wri,Banks:1996vh}. These links clearly 
warrant a sustained effort to study supersymmetric Yang-Mills theory
from all possible perspectives.

Yet, despite the huge literature on the subject, and especially the maximally 
extended $\cN=4$ theory, important questions remain. For instance, in what precise 
sense is this theory supposed to exist as a non-trivial quantum field theory beyond 
perturbation theory, and how can one ensure that it is not simply a free theory in disguise? 
The framework of Wightman axioms is not appropriate here: because of exact 
quantum conformal invariance there is no mass gap, consequently there are no 
asymptotic one particle states, and hence no $S$-matrix (at least not in any conventional sense)
whose non-triviality would affirm the non-triviality of the theory. A better framework 
is provided by the conformal bootstrap (see {\em e.g.} \cite{Rychkov:2016iqz}) where 
one must establish the existence of non-trivial correlation functions satisfying 
all the axioms of the conformal bootstrap program. 
This specifically concerns 
$n$-point correlators for $n\geq 4$ and the associated functions of the conformal cross 
ratios, whose existence beyond perturbation theory remains to be established, 
despite considerable evidence from integrability \cite{Beisert:2010jr}, progress 
with amplitude calculations \cite{Dixon:2011xs,Henn:2014yza}, and from holographic duality 
(see {\em e.g.} \cite{Alday:2021odx} and references therein).

Likewise, questions remain with regard to finiteness, especially concerning a 
non-pertur\-bative construction of the theory which would require a non-perturbative
regularization both in the IR and the UV. Beyond the vanishing of the $\beta$-function,
which has been confirmed in a variety of ways 
\cite{Sohnius:1981sn,Grisaru:1980nk,Howe:1982tm,Brink:1982pd,Howe:1983sr,Baulieu:2006ru},
the perturbative finiteness of the $\cN=4$ theory has been manifested only 
in the light-cone gauge \cite{Mandelstam:1982cb,Brink:1982wv}, whereas for 
other (in particular, covariant) gauges one has to cope with the usual quantum
field theoretic infinities (wave function renormalizations) \cite{Velizhanin:2008rw}. 
It is therefore not at all obvious how maximal supersymmetry can be usefully 
exploited towards a truly {\em non-perturbative} construction, as every non-perturbative
regularization will break supersymmetry at least partially. 

The present work is part of an ongoing effort to develop an alternative perspective 
on supersymmetric Yang-Mills theories, in order to eventually address some
of the above questions in a different way.
Our approach has its origins in one of the authors'
early work \cite{Nicolai:1980jc,Nicolai:1984jg}, according to which there exists
a non-linear and non-local transformation $\cT_g$ of the bosonic fields (Nicolai map)
which maps the full interacting functional measure to that of a free theory, and 
whose Jacobian equals the product of fermionic determinants, at least on the gauge 
hypersurface. While the expansion of $\cT_g$ for the $\cN=1$, $D=4$ theory in Landau 
gauge up to second order in the Yang-Mills coupling $g$ was already given 
in the original work, these results were only recently extended to other critical
dimensions \cite{Ananth:2020gkt} and to third order \cite{Ananth:2020lup,Ananth:2020jdr}, 
again in the Landau gauge. The latter constructions make crucial use of techniques 
developed already long ago by Dietz and Lechtenfeld 
\cite{Flume:1983sx,Dietz:1984hf,Lechtenfeld:1984me,Dietz:1985hga},
employing a certain functional integro-differential operator $\cR\equiv \cR_g$
governing the coupling constant flow. The inverse map $\cT_g^{-1}$ is then 
obtained by formally exponentiating this operator.

In this paper we extend these results in two directions. First of all 
we show how the construction generalizes to arbitrary gauges, and 
in particular to the axial gauge, which includes the light-cone gauge as a special case. 
Secondly, we present explicit formulas for the axial gauge up to second order, 
and to order $\cO(g^4)$ for the Landau gauge. These formulas illustrate 
that $\cT_g$ takes a more complicated form for gauges different from the 
Landau gauge. The privileged status of the Landau gauge follows
from general properties of the $\cR$-operator, some of which were
already discussed in \cite{Dietz:1984hf,Lechtenfeld:1984me,Dietz:1985hga},
and which will be further elaborated here. We will furthermore distinguish
between `on-shell' and `off-shell' $\cR$-operators:
this distinction goes in parallel with the usual notion of `on-shell' {\em vs.}
`off-shell' in supersymmetric theories, but here only refers to the need (or not)
to restrict the functional measure to the gauge surface. Therefore the 
`on-shell-ness' is much less of a restriction here than it is in the standard formulation of 
supersymmetric field theories: it only means that, when using the field transformation 
to perform higher order quantum computations, along the lines of \cite{Nicolai:2020tgo}, 
one must restrict the gauge parameter to the value $\xi = 0$. We note
that light-cone computations are anyway done in this way, by setting to
zero one light-cone component of the gauge field, so one can indeed ignore
the ghost determinant.

The fact that the results in the axial gauge are more complicated is in accord
with the mixed success story of the axial gauge in quantum field theory \cite{Konetschny:1975he}. 
Nevertheless, there are at least two reasons to follow up on it. 
The first is early work displaying hints of a polynomial
form of the mapping for the $\cN=1$ and $\cN=2$ theories in the light-cone gauge,
and in terms of the light-cone components of the field strength
\cite{deAlfaro:1984gw,deAlfaro:1984hb,DeAlfaro:1986uv,Lechtenfeld:1986gd}. 
Unfortunately, inspection of the relevant formulas reveals
that they do not apply to the `real' super-Yang-Mills theory. Instead, one must 
simultaneously invoke the light-cone gauge (which exists only for Lorentzian signature) 
{\em and} introduce a complexification of the basic fields, which for the fermions 
would be appropriate for Euclidean spinors. On the other hand, employing a time-like
axial gauge with Euclidean signature, a direct construction fails \cite{Nicolai:1982ye}.
A second reason comes from more recent work where it was shown that the 
maximal $\cN=4$ theory admits a reformulation where the 
Hamiltonian acquires a quadratic form in light-cone superspace
\cite{Ananth:2015tsa,Ananth:2020mws}. The relevant formulas there involve
a field re-definition in terms of the light-cone supercharge operator which likewise 
acts non-locally and non-linearly.

As the explicit formulas derived in this paper are quite involved, readers may 
wonder about their possible use. However, one should keep in mind that these 
complications are mainly due to the fact that we here consider {\em gauge-variant}
expressions (operators), something that is rarely done in more standard investigations
of $\cN=4$ super-Yang-Mills theory. If one restricts attention to gauge-invariant 
combinations, the relevant expressions simplify, because then only
the invariant part $\cR_\text{inv}$ of the $\cR$-operator contributes. We plan to return 
to these issues in future work, limiting ourselves here to a few brief comments
in the concluding section.

While finalizing the present paper we received the preprint \cite{Lechtenfeld:2021uvs} which contains very similar results, and derives an elegant formula for $\cT_g$ via a path ordered 
exponential.

\section{Pure super-Yang-Mills theories: preliminaries}
We first collect some basic and well known formulas, mainly to fix our notations 
and conventions. Throughout, we employ the `mostly minus' metric $\eta^{\mu\nu}$ with 
signature $(+,-,\cdots,-)$ and the gamma matrices $\{\c^\mu , \c^\nu\} = 2\eta^{\mu\nu}$.

\noindent
The $\cN=1$ super-Yang-Mills action $\rS_\text{inv}$ in $D=4$ dimensions 
is given by \cite{Ferrara:1974pu,deWit:1975veh}
\begin{align}\label{eq:Sinv}
{\rS}_\text{inv} [A^a_\mu,\l^a, D^a ] = \int \md x \ \left[ - \frac{1}{4} F_{\m \n}^a F^{a \, \m \n}
 - \frac{i}{2} \bar{\l}^a \c^\m (\mD_\m \l)^a + \frac{1}{2} D^a D^a \right] 
\end{align}
with the standard definitions
\begin{align}
F_{\mu\nu}^a &\coloneqq \pa_\mu A_\nu^a - \pa_\nu A_\mu^a + gf^{abc} A_\mu^b A_\nu^c \, , \\
(\mD_\mu \l)^a &\coloneqq \pa_\m \l^a + gf^{abc} A_\m^b \l^c \, , 
\end{align}
where $g$ is the coupling constant, and $f^{abc}$ are the structure constants of
the group in question (usually $\mathrm{SU}(N)$). $\l^a$ is a Majorana spinor, and $D^a$ is 
the auxiliary field which is needed to close the super-algebra off-shell. 
The action \eqref{eq:Sinv} is invariant under the supersymmetry variations
\begin{align}\label{eq:SUSYVariations}
&\d_\a A_\m^a = \left( i \bar{\l}^a \c_\m \right)_\a \, , &&\d_\a \l_\beta^a = 
-\frac{1}{2} (\c^{\m\nu})_{\beta\alpha} F_{\m \n}^a + i (\c^5)_{\beta \a} D^a \, , 
&&\d_\a D^a = \left( \mD^\m \bar{\l}^a \c^5 \c_\m\right)_\a \, , 
\end{align}
where we have stripped off the (anti-commuting) supersymmetry parameter. 
Thanks to the presence of the auxiliary field,
the supersymmetric action \eqref{eq:Sinv} can be written as a super-variation, {\em viz.}
\begin{align}\label{eq:dDelta}
\rS_\text{inv} = \d_\a \D_\a
\end{align}
with
\begin{align}\label{eq:Delta}
\D_\a \coloneqq \int \md x \ \left[ - \frac{i}{16} (\c^{\mu\nu} \l^a)_\a F^a_{\mu\nu}
 + \frac18 (\c^5 \l^a)_\a D^a \right] \, . 
\end{align}
For the full action $\rS \equiv \rS_\text{inv} + \rS_\text{gf}$ we also need the gauge fixing 
term (see {\em e.g.} \cite{Ramond:1981pw})
\begin{align}\label{eq:Sgf}
{\rS}_\text{gf}[A_\mu^a,C^a,\bar{C}^a] = \int \md x \ \left[ \frac{1}{2\xi} \, \cG^a[A] \cG^a[A] 
+ \bar{C}^a \frac{\d \cG^a[A]}{\d A_\m^b} (\mD_\m C)^b \right] 
\end{align}
with the ghost and anti-ghost fields $C^a$ and $\bar{C}^a$, and the
gauge fixing functional $\cG^a[A](x)$ with gauge parameter $\xi$ (usually 
taken as $\xi =1$, while $\xi \ra 0$ corresponds to the insertion of the 
delta functional $\prod_x \cG^a[A](x)$ in the functional measure).
The combined action $\rS$ is then invariant under the BRST (Slavnov) transformations:
\begin{align}
\begin{aligned}\label{eq:SlavnovVariations0}
&s(A_\m^a) = (\mD_\m C)^a \, , &&s(F_{\m\nu}^a) = f^{abc} F_{\m\nu}^b C^c \, , 
&&s( \l^a ) =f^{abc} \l^b C^c \, , \\
&s(C^a) = -\frac{g}{2} f^{abc} C^b C^c \, , &&s(\bar{C}^a) = - \frac{1}{\xi} \, \cG^a[A] \, 
, && s(D^a) = f^{abc} D^b C^c \, .
\end{aligned}
\end{align}
The most general gauge fixing functional compatible with our construction is any 
functional obeying the scaling relation
\begin{equation}\label{Ga0}
\cG^a [A] = g\, \cG^a[g^{-1} A]
\end{equation}
Although our derivation is thus valid for a very large class of possibly non-local and non-linear
gauge functionals, we will mostly restrict attention to linear and local gauge fixing conditions
\begin{align}\label{eq:Ga}
\cG^a[A](x) \,\equiv \, \cG^\mu A_\mu^a(x)
\end{align}
in the remainder, so that for $\cG^\mu = \pa^\mu$ and $\cG^\mu = n^\mu$, respectively, 
we recover the Landau and axial gauges (or light-cone gauge, if $n^\mu$ is null).

For the pure super-Yang-Mills theories in $D=6$ and $D=10$ dimensions 
there are no fully supersymmetric off-shell formulations (at least not with finitely many 
auxiliary fields), but the on-shell Lagrangians are the same as in \eqref{eq:Sinv}
with $D^a =0$, keeping in mind that the gauginos are Weyl, and Majorana-Weyl, 
respectively, in those dimensions \cite{Brink:1976bc}. The formula \eqref{eq:dDelta} 
can therefore not be directly applied to the extended theories in $D=6$ and $D=10$: 
without auxiliary fields, the variation of $\rS_\text{inv}$ w.r.t. the coupling constant 
produces extra terms which cannot be written as super-variations, cf. appendix A 
of \cite{Ananth:2020lup}. Nevertheless, the vacuum energy vanishes for these 
theories as well; this can be seen for instance by formulating them
in a partially off-shell version by re-writing them in terms of $\cN=1$ off-shell 
supermultiplets. Such a re-writing would actually suffice for our purposes here,
as all we need is a formulation where the action $\rS_\text{inv}$ can be expressed
as a super-variation. Although the closure of the super-algebra
is not a relevant criterion in a formulation where all fermions have been integrated out,
we will see that the distinction between `on-shell' and `off-shell' still persists,
in that the main statements of section \ref{sec:BasicProperties} below are valid only on the gauge 
surface $\cG^a[A] = 0$ for the `on-shell' $\cR$-prescription.

The derivation of the off-shell $\cR$-prescription will necessitate a `detour'
via a reformulation of the theory in terms of rescaled fields
\begin{align}
\tA_\mu^a = g A_\mu^a \;,\q \tl^a = g\l^a \;,\q\tD^a = g D^a\; , \q
 \tC^a = gC^a\;,\q \bar{\tC}^a = g \bar{C}^a \, , 
\end{align}
such that the coupling constant appears only as an overall factor outside
\begin{align}\label{eq:Sinv1}
\tilde\rS_\text{inv} [\tA^a_\mu,\tl^a, \tD^a ] = 
\frac1{g^2} \int \md x \ \left[ - \frac{1}{4} \tF_{\m \n}^a \tF^{a \, \m \n}
 - \frac{i}{2} \bar{\tl}^a \c^\m (\mD_\m \tl)^a + \frac{1}{2} \tD^a \tD^a \right] \, , 
\end{align}
where now
\begin{align}
\tF_{\mu\nu}^a & \,\equiv\, \pa_\mu \tA_\nu^a - \pa_\nu \tA_\mu^a 
+ f^{abc} \tA_\mu^b \tA_\nu^c \, , \\
\mD_\mu \tl^a &\,\equiv\, \pa_\m \tl^a + f^{abc} \tA_\m^b \tl^c \, .
\end{align}
The ghost action $\tilde\rS_\text{gf}$, the supersymmetry and the BRST transformations 
are obtained from \eqref{eq:SUSYVariations} and \eqref{eq:SlavnovVariations0}
by dropping $g$ and putting tildes on all fields; idem for \eqref{eq:dDelta} and \eqref{eq:Delta}
(it is here that we need the scaling relation (\ref{Ga0})).
For clarity of notation we always put tildes on all quantities involving rescaled fields.

\noindent
In both formulations correlation functions are given by the standard formula
\begin{align}\label{eq:Corr1}
\big\langle\!\! \big\langle X[A] 
\big\rangle\!\!\big\rangle_g = \int \cD A \ \cD\l \ \cD C \ \cD\bar{C} \ X[A] 
\ e^{-i{\rS}[g, A, \l, C,\bar{C}]} \, ,
\end{align}
where $X$ is some functional (usually a monomial) in the gauge fields;
since we do not consider matter couplings nor expectation values with 
the auxiliary $D^a$-fields, we can ignore them (and eliminate them by trivial 
Gaussian integration). The formula for the tilded fields is analogous, so that for instance
\begin{align}
\big\langle\!\! \big\langle \tA_{\m_1}^{a_1}(x_1) \ldots \tA_{\m_n}^{a_n}(x_n)
\big\rangle\!\!\big\rangle_g = g^n\,
\big\langle\!\! \big\langle A_{\m_1}^{a_1}(x_1) \ldots A_{\m_n}^{a_n}(x_n)
\big\rangle\!\!\big\rangle_g \, .
\end{align}
Either way, there is no need for a normalizing factor for the expectation
value because of the (piecewise) constancy of the vacuum functional
$\langle\!\!\langle \bE \rangle\!\!\rangle_g$ as a function of the coupling
parameters (vanishing vacuum energy in supersymmetric theories). 
As in \cite{Ananth:2020lup}, we can re-express the expectation value 
by means of a {\em purely bosonic functional integral}
\begin{align}
\big\langle X[A] \big\rangle_g = \int \cD_g[A] \ X[A] \qquad \Big( = 
\big\langle\!\! \big\langle X[A] \big\rangle\!\!\big\rangle_g \Big) \, ,
\end{align}
where the non-local functional measure $\cD_g[A]$ is obtained by integrating 
out all anti-commuting fields (gauginos and ghosts), with an analogous formula
for the rescaled fields.

There are thus two versions of the theory in which to consider the limit
$g \ra 0$. For the untilded version, the limit of
$\rS_\text{inv} + \rS_\text{gf}$ is simply the free supersymmetric Maxwell theory. 
By contrast, the $g\ra 0$ limit of $\tilde\rS_\text{inv} + \tilde\rS_\text{gf}$ localizes 
the bosonic Yang-Mills action on zero curvature configurations. Here we will be 
concerned with the former case, and make use of the tilded formulation only
as an intermediate device.

\section{The \texorpdfstring{$\ctR$}{R}-operator}
\subsection{Basic properties}\label{sec:BasicProperties}
The aim is now to construct the transformation $\cT_g$ \cite{Nicolai:1980jc,Nicolai:1984jg}
which maps the functional measure to a free measure so that
\begin{align}
\Big\langle X[A] \Big\rangle_g = \Big\langle X\big[\cT^{-1}_g[A]\big] \Big\rangle_0 =
\int \cD_0[A] \ X\big[\cT^{-1}_g[A]\big] \, .
\end{align}
More specifically, denoting bosonic Yang-Mills action by $\rS_\text{YM}[A,g]$ this
means in particular
\begin{align}\label{eq:S0}
\rS_\text{YM}\big[A,g\big] = \rS_\text{YM} \big[\cT_g[A],0\big] \, .
\end{align}
Furthermore, the map should have the property that the Jacobian of the transformation
$\cT_g$ equals the product of the fermionic determinants
\begin{align}\label{eq:det}
\det \left( \frac{\d \cT_g[A]}{\d A}\right) = \D_{\mathrm{MSS}}[A] \,\D_{\mathrm{FP}}[A] \, , 
\end{align}
at least on the gauge surface $\cG^a[A]=0$.
Here the Matthews-Salam-Seiler determinant $\D_{\mathrm{MSS}}[A]$ 
\cite{Matthews:1954zg,Seiler:1974ne} is obtained by integrating 
out the gauginos\footnote{Because $\l^a$ is Majorana, $\D_{\mathrm{MSS}}$ is really
 a Pfaffian.}, and $\D_{\mathrm{FP}}[A]$ is the Faddeev-Popov determinant 
 \cite{Faddeev:1967fc,tHooft:1971akt}.
The map $\cT_g$ is constructed iteratively in terms of a generating functional
differential operator $\cR_g$, such that
\begin{align}\label{eq:InverseNicolaiMap}
\left(\cT_g^{-1} A\right)_\m^a(x) = \sum_{n=0}^\infty \frac{g^n}{n!}
 \left((\cR_g^n A)_\m^a(x) \, \bigg\vert_{g=0}\right) \, .
\end{align}
The operator $\cR_g$ is determined from the flow equation 
\cite{Dietz:1984hf,Lechtenfeld:1984me}
\begin{align}
\frac{\md}{\md g} \big\langle X \big\rangle_g = \big\langle \cR_g (X) \big\rangle_g
\end{align}
and should act distributively:
\begin{align}
\cR_g (XY) = \cR_g(X) Y + X\cR_g(Y) \, .
\end{align}
Furthermore the statement \eqref{eq:S0} is equivalent to
\begin{align}
\cR_g \left( \int \md x \ F_{\m \n}^a F^{a \, \m \n} \right) = 0 \, .
\end{align}
Finally we require
\begin{align}
\cR_g \left( \cG^a[A]\right) = 0 \, .
\end{align}
Below we will perform this construction in the tilded formulation and
compare the resulting expression for the new $\ctR_g$-operator with the $\cR_g$-operator
obtained in \cite{Ananth:2020lup}. Importantly, in the limit $g\ra 0$ these operators differ 
by terms involving the Landau gauge condition, and in general the latter do not vanish on the
gauge surface $\cG^a[A] =0$ if $\cG^a$ is different from the Landau gauge.
Consequently, even though we are ultimately interested in constructing
$\cT_g$ in the untilded formulation it turns out that for gauges other than
the Landau gauge we have to perform the construction first for the tilded
version, because it reveals the existence of terms that cannot be obtained from the
on-shell $\cR$-prescription.

\subsection{On-shell \texorpdfstring{$\cR$}{R}-operator}
Building on earlier results of \cite{Flume:1983sx}, it was shown
in \cite{Ananth:2020lup} that for all pure super-Yang-Mills theories
in dimensions $D=3$, $4$, $6$ and $10$, and with the Landau gauge, the $\cR$-operator 
can be represented in the form
\begin{align}\label{eq:RFinal}
\cR_g= \cR_\text{inv} + \cR_\text{gf}
\end{align}
with
\begin{align}\label{eq:RInvFinal}
\cR_\text{inv} \coloneqq \frac{\md}{\md g} - \frac{1}{2r} \int \md x \ \md y \ \tr \left( \c_\m S^{ab}(x,y;A) \c^{\r \l} \right) f^{bcd} A_\r^c(y) A_\l^d(y) \, \frac{\d}{\d A_\m^a(x)}
\end{align}
and 
\begin{align}
\begin{gathered}\label{eq:RGfFinal}
\cR_\text{gf} \coloneqq \\
- \frac{1}{2r} \int \md x \ \md y \ \md z \ (D_\m G)^{ae}(x,z;A) \tr \left( \c^\n \pa_\n S^{eb}(z,y;A) \c^{\r\l} \right) f^{bcd} A_\r^c(y) A_\l^d(y) \, \frac{\d}{\d A_\m^a(x)} \, ,
\end{gathered}
\end{align}
where $r=2(D-2)$ is the number of effective gaugino degrees of freedom.
The gaugino and ghost propagators in the gauge field background
given by $A_\mu^a(x)$ appearing in these expressions are defined by
\begin{align}
\c^\m (\mD_\m S)^{ab}(x,y; A) &= \d^{ab} \d(x-y) \, , \\
\frac{\d \cG^a}{\d A_\mu^c} (\mD_\mu G)^{cb}(x,y;A) & \equiv 
\pa^\mu (\mD_\mu G)^{ab}(x,y;A) = \d^{ab} \d(x-y) \, .
\end{align}
For practical calculations it is sometimes useful to write out these equations
in Dyson-Schwinger (integrated) form
\begin{align} \label{eq:DS}
\begin{aligned}
S^{ab}(x,y;A) &= \d^{ab} S_0(x-y) - g f^{acd} \int \md z \ S_0(x-z) \c^\mu A_\mu^c(z) S^{db}(z,y;A) \, , \\
G^{ab}(x,y;A) &= \d^{ab} G_0(x-y) - g f^{acd} \int \md z \ G_0(x-z) \cG^\mu A_\mu^c(z) G^{db}(z,y;A) 
\end{aligned} 
\end{align}
for linear gauge functions of the form \eqref{eq:Ga}.

\noindent
While the above prescription works for all pure super-Yang-Mills theories, 
it is subject to the following restrictions \cite{Ananth:2020lup}:
\begin{itemize}
\item It only works for the Landau gauge $\cG^a[A] \equiv \pa^\m A_\m^a$.
\item $\cR_g$ acts distributively only on the gauge surface $\pa^\m A_\m^a = 0$,
 corresponding to the limit $\xi\ra 0$ in $\rS_\text{gf}$ where the measure contains
 the delta functional $\prod_x \d\big(\pa^\mu A_\mu^a(x)\big)$.
\item Beyond order $\cO(g^2)$ the equality \eqref{eq:det} of the functional 
 Jacobian and the product of fermionic determinants likewise holds 
 only on the gauge surface $\pa^\m A_\m^a = 0$.
\end{itemize}
In particular, the prescription does {\em not} work for the axial and light-cone gauges, 
for which one encounters discrepancies in the construction of $\cT_g$ already 
at order $\cO(g^2)$.

\subsection{\texorpdfstring{$\ctR$}{R}-operator for rescaled fields}
Now, already in 1984 Dietz and Lechtenfeld constructed a $\ctR$-operator for 
the rescaled (tilded) $\cN=1$, $D=4$ theory, and for general $\cG^a[\tA]$
\cite{Dietz:1984hf,Lechtenfeld:1984me,Dietz:1985hga}. With our notation 
and conventions, their result for $\ctR_g = \ctR_\text{inv} + \ctR_\text{gf}$ reads
\begin{align}\label{eq:RInvFinalR}
\begin{gathered}
\ctR_\text{inv} = \frac{\md}{\md g} + \frac{1}{8g} \int \md x \ \md y \ \tr \big( \c_\m 
\tS^{ab}(x,y;\tA) \c^{\r \l} \big) \tF_{\r\l}^b(y) \, \frac{\d}{\d \tA_\m^a(x)} + \frac1{g} \int \md x \ \tD^a(x) \, \frac{\d}{\d \tD^a(x)}   \\
+ \frac{i}{4g} \int \md x \ \tr \big( \c_5 \c_\m \tS^{ab}(x,y;\tA) \big) \tD^b(x) \, \frac{\d}{\d \tA_\m^a(x)} 
\end{gathered}
\end{align}
and
\begin{align}\label{eq:RGfFinalR}
\begin{aligned}
\ctR_\text{gf} &= - g \int \md x \ \md y \ (\mD_\m \tG)^{ab}(x,y;\tA) \, \ctR_\text{inv}\left( \textstyle\frac{1}{g} \, \cG^b[\tA](y) \right) \, \frac{\d}{\d \tA_\m^a(x)}  \\
&\quad + g f^{abc} \int \md x \ \md y \  \tG^{bd}(x,y;\tA) \, \ctR_\text{inv}\left( \textstyle\frac{1}{g} \, \cG^d[\tA](y) \right) \, \tD^c(x) \, \frac{\d}{\d \tD^a(x)} 
\end{aligned}
\end{align}
(now with $r=4$). For the reader's convenience we summarize
the derivation of this result in appendix \ref{sec:AppendixConstruction}.
In the remainder we will disregard all terms involving the auxiliary fields $\tD^a$
as the relevant expressions considered here do not depend on them.

While $\cR$ is specific to the Landau gauge, but works for all
critical dimensions, $\ctR$ exists for any gauge functional and is manifestly 
distributive\footnote{As we pointed out, analogous off-shell $\ctR$-operators 
 can in principle be constructed for the extended theories in $D=4$ by formulating them
 in terms of off-shell $\cN=1$ supermultiplets.}.
The corresponding inverse map is obtained as
\begin{align}\label{eq:InverseNicolaiMapR}
\left( \cT_g^{-1} (A)\right)_\mu^a \,\equiv \,
\left( \ctT_g^{-1} \left( \textstyle\frac{1}{g} \tA \right) \right)_\m^a(x) \coloneqq 
\sum_{n=0}^\infty \frac{g^n}{n!} \left[ \left( \ctR_{g}^n \left( \textstyle\frac{1}{g} \tA \right) \right)_\m^a(x) 
\, \bigg\vert_{\tA= gA} \bigg\vert_{g=0} \right] \, .
\end{align}
The question which we wish to address here is how the operators $\cR$ and $\ctR$ 
are precisely related. The main step will be the demonstration that a proper 
limit $g\ra 0$ exists also for $\ctR_g$, which should then yield the above prescription.
To compare the two prescriptions we first rewrite \eqref{eq:RInvFinalR} and \eqref{eq:RGfFinalR}
by means of the identity \cite{Lechtenfeld:1984me}
\begin{align}
\c^{\r\l} \tF_{\r\l}^b = 2 \c^\r \c^\l (\mD_\r \tA_\l)^b - 2 \pa^\l \tA_\l^b - 
f^{bde} \c^{\r\l} \tA_\r^d \tA_\l^e \, ,
\end{align}
leaving the gauge functional \eqref{eq:Ga} arbitrary. Integrating by 
parts, so $\mD_\rho$ acts on the fermionic propagator to give a $\d$-function, 
then leads to the new representation
\begin{align}
\ctR_g = \ctR_0 + \ctR_1 + \ctR_2
\end{align}
with the counting operator (now with $\tD^a=0$)
\begin{align}\label{eq:CountingOp}
\ctR_0 \coloneqq \frac{\md}{\md g} + \frac{1}{g} \int \md x \ \tA_\mu^a(x) \, \frac{\delta}{\delta \tA_\mu^a(x)} \, .
\end{align}
The other two operators are given by
\begin{align}
\begin{gathered}\label{eq:tR}
\ctR_1 \coloneqq - \frac{1}{8g} \int \md x \ \md y \ \tr \left( \c_\m \tS^{ab}(x,y;\tA) \c^{\r \l} \right) f^{bcd} \tA_\r^c(y) \tA_\l^d(y) \frac{\d}{\d \tA_\m^a(x)} \\
- \frac{1}{8g} \int \md x \ \md y \ \md z \ (\mD_\m \tG)^{ae}(x,z;\tA)
 \tr \left( \c_\n \frac{\d\cG^e}{\d \tA_\nu^f} \tS^{fb}(z,y;\tA) \c^{\r\l} \right) 
 f^{bcd} \tA_\r^c(y) \tA_\l^d(y) \, \frac{\d}{\d \tA_\m^a(x)} 
\end{gathered}
\end{align}
and 
\begin{align}
\begin{gathered}\label{eq:RestTerm}
\ctR_2 \coloneqq - \frac{1}{4g} \int \md x \ \md y \ \tr \left( \c_\mu \tilde{S}^{ab}(x,y;\tA) \right) \pa^\l \tA_\l^b(y) \, \frac{\d}{\d \tA_\m^a(x)} \\
+ \frac{1}{4g} \int \md x \ \md y \ \md z \ (\mD_\m \tG)^{ab}(x,y;\tA) 
 \tr \left( \c_\n \frac{\d \cG^b}{\d \tA_\nu^e} 
 \tS^{ec}(y,z;\tA) \right) \pa^\l \tA_\l^c(z) \, \frac{\d}{\d \tA_\m^a(x)} \, .
\end{gathered}
\end{align}
The counting operator $\ctR_0$ obeys
\begin{align}
\ctR_0 (A_\mu^a) \equiv \ctR_0\left( \frac1{g}\tA_\mu^a \right) = 0
\end{align}
as well as relations like
\begin{align}
\ctR_0 \left( \tS^{ab}(x,y;\tA) \right) = -  f^{cde}  \int \md z \ \tS^{ac}(x,z;\tA)
 \c^\mu  \tA_\mu^d(z) \tS^{eb}(z,y;\tA)
\qquad etc.
\end{align}
It is readily seen that $\ctR_1$ coincides with the relevant terms from
\eqref{eq:RInvFinal} and \eqref{eq:RGfFinal} for $r=4$ upon substituting 
$\tA_\mu^a = g A_\mu^a$ and adopting the Landau gauge. 
By contrast, the new term $\ctR_2$ has no analog in the on-shell $\cR$-operator,
as it vanishes for $\pa^\l \tA_\l^a =0$. However, off the gauge surface
it does contribute and thus contains relevant information even for the Landau gauge.
In evaluating it, one must first show that it possesses a well defined limit 
for arbitrary gauge functionals $\cG^a[\tA]$ subject to the condition (\ref{Ga0})
upon setting $\tA_\mu^a = g A_\mu^a$ 
and taking $g\ra 0$. To prove this we need to consider the potentially singular
zeroth order contributions in both integrands of \eqref{eq:RestTerm}, using \eqref{eq:DS}, 
\begin{align}
\tS^{ab} (x,y;\tA) &= - \d^{ab} \c^\rho \pa_\rho C(x-y) + \cO(\tA) \, , \\
\tG^{ab} (x,y;\tA) & = \d^{ab} \tG_0(x-y) + \cO(\tA) \, , \\
\frac{\d\cG^a[\tA](x)}{\d \tA^b_\mu (y)} &= \d^{ab} \cG^\mu \d(x-y) + \cO(\tA) \, , 
\end{align} 
where $C(x)$ is the free scalar propagator obeying $\Box C(x) = - \d(x)$.
By (\ref{Ga0}) we can ignore the $\cO(\tA)$ terms since they are non-singular 
as $g\ra 0$. For the Landau 
gauge ($\cG^\mu = \pa^\mu$) the cancellation of the singular term follows easily
upon use of $\c^\mu \pa_\mu S_0(x) = \d(x)$ and $G_0(x) = -C(x)$. For the axial gauge 
($\cG^\mu = n^\mu$), we compute
\begin{align}
\tr \left( \c_\mu n^\mu S_0(y-z) \right) = - 4 n^\mu \pa_\mu C(y-z) \, ,
\end{align}
integrate by parts, and use the defining equation for the free ghost propagator 
$n^\mu \pa_\mu \tG_0(x) = \d(x)$ to show that these contributions cancel again
(as we pointed out, higher order terms in the gauge functional do not affect this
argument). All remaining terms in \eqref{eq:RestTerm}
are at least of order $\tA$ and therefore possess a well-defined limit for $g\to 0$. 

The calculation of $\cR(A)$ then proceeds by first computing
$\ctR( g^{-1} \tA)$ and then expanding in $\tA$, setting $\tA= gA$. To compute the
the Taylor coefficients in \eqref{eq:InverseNicolaiMapR} we finally take the limit
$g\ra 0$. This limit yields extra contributions over and above 
the ones from the $\cR$-prescription \eqref{eq:RInvFinal} and \eqref{eq:RGfFinal}
even for the Landau gauge. For the latter these are the terms that for $D=4$ 
ensure that the equality of the determinants \eqref{eq:det} remains valid to
any order {\em even without imposing the gauge condition}.
For the axial gauge we also find extra terms, and moreover ones which do 
{\em not} vanish on the gauge hypersurface $n^\mu A_\mu^a = 0$, {\em i.e.}
\begin{align}\label{eq:limR2} 
\lim\limits_{g\to 0} \ \ctR_2 \left( \frac1{g} \tA_\mu^a \right) \bigg\vert_{n\cdot A^a = 0} \neq 0 \, .
\end{align}
This explains why the on-shell $\cR$-prescription does not work for the axial gauge: to get the
correct answer, we need to add the extra terms resulting from \eqref{eq:limR2}.

\section{The map in axial gauge}
We next apply the above prescription to determine the expansion of $\cT_g$ to second order 
for the axial gauge. By the above construction this result will contain a part identical to the
result in the Landau gauge, as well as extra terms resulting from \eqref{eq:limR2}. We shall
then verify all requisite properties. Although we start from the $\cN=1$ theory, it turns out
that at least to second order this expansion remains valid for the other critical dimensions, 
and even off the gauge surface $n^\mu A_\mu^a =0$. This is a feature which for the
extended theories we do not expect to persist in higher orders, as it would require 
a formulation of these theories at least in terms of $\cN=1$ off-shell multiplets.

\newpage
\subsection{Expansion to \texorpdfstring{$\cO(g^2)$}{O(g2)}}
\begin{footnotesize}\label{eq:Taxial}
\begin{align}
\begin{aligned}
\left(\cT_g A \right)_\m^a(x) &= A_\m^a(x) + g f^{abc} \int \md y \ \md z \ \left( \eta_{\m\n} \d(x-y) - \pa_\m G_0(x-y) n_\n \right) \\
&\qqq \times \left\{ A^{b \, \n}(y) C(y-z) \, \pa \cdot \! A^c(z) + \pa^\l C(y-z) A^{b \, \n}(z) A_\l^c(z) \right\} \\
&\q + 2 g f^{abc} \int \md y \ \md z \ \md w \ \left( \eta_{\m\n} \d(x-y) - \pa_\m G_0(x-y) n_\n \right) \\
&\qqq \times \pa_\l C(y-z) A^{b \, [\n}(z) \pa^{\l]} C(z-w) \, \pa \cdot \! A^c(w) \\
&\q + \frac{g^2}{2} f^{abc} f^{bde} \int \md y \ \md z \ \md w \ \left( \eta_{\m\n} \d(x-y) - \pa_\m G_0(x-y) n_\n \right) \ \bigg\{ \\
&\qqq - 2 A^{c \, \n}(y) C(y-z) A_\l^d(z) \pa^\l C(z-w) \, \pa \cdot \! A^e(w) \\
&\qqq - A^{c \, \n}(y) C(y-z) \, \pa \cdot \! A^d(z) C(z-w) \, \pa \cdot \! A^e(w) \\ 
&\qqq - \frac{1}{2} C(y-z) \, \pa \cdot \! A^c(z) \pa^\l C(z-w) A^{d \, \n}(w) A_\l^e(w) \\ 
&\qqq + \frac{1}{2} C(y-z) \, \pa \cdot \! A^c(z) \pa^\l C(y-w) A^{d \, \n}(w) A_\l^e(w) \\ 
&\qqq - \frac{1}{2} C(y-z) A^{d \, \n}(z) A_\l^e(z) \pa^\l C(z-w) \, \pa \cdot \! A^c(w) \\ 
&\qqq + \frac{1}{2} \pa^\l C(y-z) A^{d \, \n}(z) A_\l^e(z) C(z-w) \, \pa \cdot \! A^c(w) \\
&\qqq - 2\pa_\l C(y-z) A^{c \, [\n}(z) A^{d\, \l]}(z) C(z-w) \, \pa \cdot \! A^e(w) \\
&\qqq + 3 \pa_\rho C(y-z) A_\l^c(z) \pa^{[\n} C(z-w) A^{d\, \l}(w) A^{e \, \rho]}(w) \bigg\} \\ 
&\q + \frac{g^2}{2} f^{abc} f^{bde} \int \md y \ \md z \ \md w \ \md v \ \left( \eta_{\m\n} \d(x-y) - \pa_\m G_0(x-y) n_\n \right) \ \bigg\{ \\
&\qqq - C(y-z) A^{d \, [\n}(z) \pa^{\l]} C(z-w) \, \pa \cdot \! A^e(w) \pa_\l C(z-v) \, \pa \cdot \! A^c(v) \\ 
&\qqq - C(y-z) \, \pa \cdot \! A^c(z) \pa_\l C(z-w) A^{d\, [\n}(w) \pa^{\l]} C(w-v) \, \pa \cdot \! A^e(v) \\
&\qqq + C(y-z) \, \pa \cdot \! A^c(z) \pa_\l C(y-w) A^{d\, [\n}(w) \pa^{\l]} C(w-v) \, \pa \cdot \! A^e(v) \\
&\qqq - \pa_\l C(y-z) A^{d \, [\n}(z) \pa^{\l]} C(z-w) \, \pa \cdot \! A^e(w) C(z-v) \, \pa \cdot \! A^c(v) \\ 
&\qqq - \pa^\l C(y-z) \pa^\n C(z-w) \, \pa \cdot \! A^c(w) \pa^\rho C(z-v) A_\l^d(v) A_\rho^e(v) \\ 
&\qqq + 2 \pa_\l C(y-z) \pa^{[\n} A^{d\, \l]}(z) C(z-w) \, \pa \cdot \! A^e(w) C(z-v) \, \pa \cdot \! A^c(v) \\
&\qqq - 2 \pa_\l C(y-z) A^{c \, [\n}(z) \pa^{\l]} C(z-w) \, \pa \cdot \! A^d(w) C(w-v) \, \pa \cdot \! A^e(v) \\ 
&\qqq - 4 \pa_\l C(y-z) A^{c \, [\n}(z) \pa^{\l]} C(z-w) A_\rho^d(w) \pa^\rho C(w-v) \, \pa \cdot \! A^e(v) \\ 
&\qqq + 6 \pa_\rho C(y-z) A_\l^c(z) \pa^{[\n} C(z-w) A^{d \, \l}(w) \pa^{\rho]} C(w-v) \, \pa \cdot \! A^e(v) \bigg\} \\ 
&\q - g^2 f^{abc} f^{bde} \int \md y \ \md z \ \md w \ \md v \ \md u \ \left( \eta_{\m\n} \d(x-y) - \pa_\m G_0(x-y) n_\n \right) \\
&\qqq \times \pa^\l C(y-z) \pa^\n C(z-w) \, \pa \cdot \! A^c(w) \pa^\rho C(z-v) A_{[\l}^d(v) \pa_{\rho]} C(v-u) \, \pa \cdot \! A^e(u) \\
&\q + \cO(g^3) \, . 
\end{aligned}
\end{align}
\end{footnotesize}
Recall that $C(x)$ is the free (massless) scalar propagator, while (now with $n^\mu n_\mu =1$)
\begin{align}
G_0(x) = \varepsilon(n \,  \cdot \!x) \, \d^{(3)}(x^\perp) \,=\, - G_0(-x)
\end{align}
is the free ghost propagator for the axial gauge, with the anti-symmetric step function
$\varepsilon(x) \coloneqq \Theta(x) - \frac12$ and the
transverse coordinate $x^\perp_\mu \equiv x_\mu - n_\mu (n \,  \cdot \!x)$. In writing 
the above result we have regrouped terms in such a way that they all appear 
with the axial projector
\begin{align}\label{eq:Piax}
\Pi_{\mu\nu}(x) \coloneqq \eta_{\mu\nu} \d(x) - \pa_\m G_0(x) n_\n
\end{align}
in front. This projector obeys $n^\mu \Pi_{\mu\nu}(x) = 0$ (but $\Pi_{\mu\nu}(x) n^\nu \neq 0$!).
By the definition of the free ghost propagator $G_0$ we also have
\begin{align}\label{eq:Piaxpa}
\int \md y \ \Pi_{\mu\nu}(x-y) \pa^\nu F(y) = 0 
\end{align}
for any function $F$. Hence the second order result in axial gauge can be written in such a way 
that it differs from the off-shell result for the Landau gauge only by the insertion of this projector, 
since all terms of type \eqref{eq:Piaxpa} drop out.

Finally we point out that the above derivation is in principle valid for all $n^\mu$, regardless
whether they are time-like, space-like or null. It therefore applies 
to the light-cone gauge as well.

\subsection{Tests}
To check the above result, we now go through all relevant tests for $A_\m^{\prime \, a} 
\equiv \left(\cT_g A\right)_\m^a$. The first test
(preservation of the gauge function) 
\begin{align}\label{eq:Test0}
n^\mu A_\m^{\prime \, a}(x) = n^\mu A_\m^a(x) 
\end{align}
is trivially satisfied up to the order considered, by the defining property of the axial 
projector \eqref{eq:Piax} and the fact that it appears in front of all terms.

\subsubsection*{Free action}
Next we test the free action. By \eqref{eq:Test0} we can ignore the piece $\propto (n^\mu A_\mu^a)^2$ 
that needs to be included in the action \eqref{eq:Sgf}. Thus we need only show that
\begin{align}
\frac{1}{2} \int \md x \ A_\m^{\prime \, a}(x) (- \Box \eta^{\m\n} + \pa^\m \pa^\n ) A_\n^{\prime a}(x) = \frac{1}{4} \int \md x \ F_{\m\n}^a(x) F^{a\, \m\n}(x) + \cO(g^3) \, .
\end{align}
We notice that any term which can be written as $\pa_\m^x \left( \ldots \right)$ does 
not contribute by the gauge invariance of the free action.
At first order we find
\begin{align}
\begin{aligned}
&\frac{1}{2} \int \md x \ A_\m^{\prime \, a}(x) (- \Box \eta^{\m\n} + \pa^\m \pa^\n ) A_\n^{\prime \, a}(x) \bigg\vert_{\cO(g^1)} \\
&\q = f^{abc} \int \md x \ \md y \ \left\{ A_\m^b(x) C(x-y) \, \pa \cdot \! A^c(y) + \pa^\l C(x-y) A_\m^b(y) A_\l^c(y) \right\} \\
&\qq \q \times (- \Box \eta^{\m\n} + \pa^\m \pa^\n ) A_\n^{\prime a}(x) \\
&\qq + 2 f^{abc} \int \md x \ \md y \ \md z \ \pa^\l C(x-y) A_{[\m}^b(y) \pa_{\l]} C(y-z) \, \pa \cdot \! A^c(z) \\
&\qq \q \times (- \Box \eta^{\m\n} + \pa^\m \pa^\n ) A_\n^{\prime a}(x) \, .
\end{aligned}
\end{align}
We integrate by parts and remove anti-symmetric terms
\begin{align}
\begin{aligned}
&\frac{1}{2} \int \md x \ A_\m^{\prime \, a}(x) (- \Box \eta^{\m\n} + \pa^\m \pa^\n ) A_\n^{\prime \, a}(x) \bigg\vert_{\cO(g^1)} \\
&\q = - f^{abc} \int \md x \ \pa_\l A_\m^a(x) A^{b \, \m}(x) A^{c \, \l}(x) \\
&\qq + f^{abc} \int \md x \ \md y \ \bigg\{ - \Box A_\m^a(x) A^{b \, \m}(x) C(x-y) \, \pa \cdot \! A^c(y) \\
&\qqq \q + \pa_\m \, \pa \cdot \! A^a(x) A^{b \, \m}(x) C(x-y) \, \pa \cdot \! A^c(y) - 2 \pa_\l A_\m^a(x) A^{b \, [\m}(x) \pa^{\l]} C(x-y) \, \pa \cdot \! A^c(y) \bigg\} \\
&\q = f^{abc} \int \md x \ \pa_\m A_\l^a(x) A^{b \, \m}(x) A^{c \, \l}(x) = \frac{1}{4} \int \md x \ F_{\m\n}^a(x) F^{a \, \m\n}(x) \bigg\vert_{\cO(g^1)} \, .
\end{aligned}
\end{align}
At the second order the steps are generally the same. Again we can disregard half the terms because of the axial projector $\Pi_{\mu\nu}(x)$. Performing similar partial integrations as above yields
\begin{align}
\begin{aligned}
&\frac{1}{2} \int \md x \ A_\m^{\prime \, a}(x) (- \Box \eta^{\m\n} + \pa^\m \pa^\n ) A_\n^{\prime \, a}(x) \bigg\vert_{\cO(g^2)} \\
&\q= \int \md x \ A_\m^{\prime \, a}(x) \big\vert_{\cO(g^2)} ( - \Box \eta^{\m\n} + \pa^\m \pa^\n ) A_\n^{\prime a}(x) \big\vert_{\cO(g^0)} \\
&\qq + \frac{1}{2} \int \md x \ A_\m^{\prime \, a}(x) \big\vert_{\cO(g^1)} ( - \Box \eta^{\m\n} + \pa^\m \pa^\n ) A_\n^{\prime a}(x) \big\vert_{\cO(g^1)} \\
&\q= - \frac{g^2}{4} f^{abc} f^{bde} \int \md x \ A_\m^a(x) A_\l^c(x) A^{d\, \m}(x) A^{e \, \l}(x) \\ 
&\qq - \frac{g^2}{2} \int \md x \ \md y \ A_\m^a(x) A_\l^e(x) \pa^\l A^{d \, \m}(x) C(x-y) \, \pa \cdot \! A^c(y) \\
&\qqq \q \times \left( f^{abc} f^{bde} + f^{eba} f^{bdc} + f^{cbe} f^{bda} \right) \\
&\q= - \frac{g^2}{4} f^{abc} f^{bde} \int \md x \ A_\m^a(x) A_\l^c(x) A^{d\, \m}(x) A^{e \, \l}(x) = \frac{1}{4} \int \md x \ F_{\m\n}^a(x) F^{a \, \m\n}(x) \bigg\vert_{\cO(g^2)} \, ,
\end{aligned}
\end{align}
where, in the second to last step, we have used the Jacobi identity
\begin{align}
\begin{aligned}
0 = f^{abc} f^{bde} + f^{eba} f^{bdc} + f^{cbe} f^{bda} \, .
\end{aligned}
\end{align}
These relations are independent of dimension.

\subsubsection*{Jacobians, fermion and ghost determinants}
Finally we need to perturbatively show that the Jacobian determinant is equal to the product of 
the MSS and FP determinants. This is done as usual order by order in $g$ by considering the 
logarithms of the determinants, {\emph i.e.}
\begin{align}\label{eq:Determinants}
\log \det \left( \frac{\d A_\m^{\prime \, a}(x)}{\d A_\n^b(y)} \right) \overset{!}{=} \log \left( \D_\text{MSS}[A] \D_\text{FP}[A] \right) \, .
\end{align}
As it turns out, for \eqref{eq:Taxial} this equality is actually valid for all critical dimensions
and off the gauge surface $n^\m A_\m^a=0$ up to the order considered (but
we do not expect this feature to persist in higher orders). For this reason we 
re-instate the general values $r$ and $D$ in the formulas below.

\noindent
The ghost determinant is computed from the functional matrix
\begin{align}
\bX^{ab}(x,y;A) = g f^{abc} G_0(x-y) \, n \! \cdot \! A^c(y) \, , 
\end{align}
using the well-known equation 
\begin{align}\label{eq:logdet}
\log \det (1 - \bX) = \tr \log (1 - \bX) \, .
\end{align}
Up to $\mathcal{O}(g^2)$ this yields
\begin{align}
\log \,\det (1 - \bX ) = \frac{g^2}{2} N \int \md x \ \md y \ G_0(x-y) \, n \!\cdot\! A^a(y) G_0(y-x) \,
n \!\cdot\! A^a(x) + \mathcal{O}(g^3) \, ,
\end{align}
where we used $f^{abc} f^{abd} = N \d^{cd}$. The relevant kernel for the MSS determinant is
\begin{align}
\bY_{\a\b}^{ab}(x,y;A) = g f^{abc} \pa_\r C(x-y) \left( \c^\r \c^\l \right)_{\a\b} A_\l^c(y) \, .
\end{align}
Because of the Majorana condition we must include an extra factor of $\textstyle\frac12$ in the expansion \eqref{eq:logdet} and get
\begin{align}
\begin{aligned}\label{eq:MSS}
\frac{1}{2} \log\det(1- \bY) &= \frac{g^2}{4} N \, \tr \left( \c^\r \c^\l \c^\sigma \c^\n \right) \int \md x \ \md y \ \pa_\r C(x-y) A_\l^a(y) \pa_\sigma C(y-x) A_\n^a(x) \\
&\quad + \mathcal{O}(g^3) \, .
\end{aligned}
\end{align}
For both determinants there is no contribution at $\mathcal{O}(g^1)$ and also there is no contribution from the Jacobi determinant at this order. Taking the trace in \eqref{eq:MSS} and multiplying the two determinants yields the right hand side of \eqref{eq:Determinants}
\begin{align}
\begin{aligned}\label{eq:FPMSSDet}
\log \left( \D_\text{MSS}[A]\, \D_\text{FP}[A] \right) \big\vert_{\mathcal{O}(g^2)} &= \frac{g^2}{2} N \int \md x \ \md y \ \bigg\{ \\
&\qqq+ r \, \pa^\m C(x-y) A_\m^a(y) \pa^\n C(y-x) A_\n^a(x) \\
&\qqq- \frac{r}{2} \, \pa_\m C(x-y) A_\n^a(y) \pa^\m C(y-x) A^{a\, \n}(x) \\
&\qqq+ G_0(x-y) \, n \,  \cdot \! A^a(y) G_0(y-x) \, n \! \cdot \! A^a(x) \bigg\} \, .
\end{aligned}
\end{align}
At $\mathcal{O}(g^2)$ the logarithm of the Jacobi determinant consists of two terms
\begin{align}
\log \det \left( \frac{\d A_\m^{\prime \, a}(x)}{\d A_\n^b(y)} \right) \bigg\vert_{\mathcal{O}(g^2)} = \tr \left[ \frac{\d A^\prime}{\d A} \bigg\vert_{\mathcal{O}(g^2)} \right] - \frac{1}{2} \tr \left[ \frac{\d A^\prime}{\d A} \bigg\vert_{\mathcal{O}(g^1)} \frac{\d A^\prime}{\d A} \bigg\vert_{\mathcal{O}(g^1)} \right]
\end{align}
and the final trace is done by setting $\m=\n$, $a=b$, $x=y$ and integrating over $x$. The computation is straightforward but we must be careful with formally divergent terms. 
Subsequently we find
\begin{align}
\begin{aligned}\label{eq:Jac1}
- \frac{1}{2} \tr \left[ \frac{\d A^\prime}{\d A} \bigg\vert_{\mathcal{O}(g^1)} \frac{\d A^\prime}{\d A} \bigg\vert_{\mathcal{O}(g^1)} \right] &= N g^2 \int \md x \ \md y \ \bigg\{ \\
&\qq+ \frac{D}{2} \pa^\m C(x-y) A_\m^a(y) \pa^\n C(x-y) A_\n^a(x) \\
&\qq+ \frac{1}{2} G_0(x-y) \, n \! \cdot \! A^a(y) G_0(y-x) \, n \! \cdot \! A^a(x) \\ 
&\qq \color{green!20!blue} + C(x-y)  \, \pa^\m \left( A_\m^a(y) G_0(y-x) \right)  \, n \! \cdot \! A^a(x)  \\ 
&\qq \color{green!20!blue} + \frac{2-D}{8} C(x-y) \, \pa \cdot \! A^a(y) \left( C(y-x) - 2 C(0) \right) \, \pa \cdot \! A^a(x) \color{black}  \bigg\} \\ 
&\q+ N g^2 \int \md x \ \md y \ \md z \ \bigg\{ \\
&\qq \color{green!20!blue} - \frac{1}{4} G_0(x-z) n^\m C(z-x) \, \pa \cdot \! A^a(y)  \pa_\m C(y-x) \, \pa \cdot \! A^a(x) \\ 
&\qq \color{green!20!blue} - 2 G_0(x-z) \pa^\m C(z-x) \, \pa \cdot \! A^a(y) n^\n \pa_{ \{\n} C(y-x) A_{\m \}}^a(x) \\
&\qq \color{green!20!blue} + 2 G_0(x-z) \pa^\m C(z-y) A_\n^a(y) n^\l \pa^\n \pa_{\{\l} C(y-x) A_{\m\}}^a(x) \\
&\qq \color{green!20!blue} + \frac{1-D}{2} \d(0) C(z-y) \, \pa \cdot \! A^a(y) C(z-x) \, \pa \cdot \! A^a(x)  \color{black} \bigg\} \, .
\end{aligned}
\end{align}
The other term gives
\begin{align}
\begin{aligned}\label{eq:Jac2}
\tr \left[ \frac{\d A^\prime}{\d A} \bigg\vert_{\mathcal{O}(g^2)} \right] &= N g^2 \int \md x \ \md y \ \bigg\{ \\ 
&\qq - \frac{4-D}{2} \pa^\m C(x-y) A_\m^a(y) \pa^\n C(y-x) A_\n^a(x) \\
&\qq + \frac{2 - D}{2} \pa_\m C(x-y) A_\n^a(y) \pa^\m C(y-x) A^{a \, \n}(x) \\ 
&\qq \color{green!20!blue} - C(x-y)  \, \pa^\m \left( A_\m^a(y) G_0(y-x) \right)  \, n \! \cdot \! A^a(x)  \\ 
&\qq \color{green!20!blue} - \frac{2-D}{8} C(x-y) \, \pa \cdot \! A^a(y) \left( C(y-x) - 2 C(0) \right) \, \pa \cdot \! A^a(x)  \color{black} \bigg\} \\ 
&\q+ N g^2 \int \md x \ \md y \ \md z \ \bigg\{ \\
&\qq \color{green!20!blue} + \frac{1}{4} G_0(x-z) n^\m C(z-x) \, \pa \cdot \! A^a(y)  \pa_\m C(y-x) \, \pa \cdot \! A^a(x) \\ 
&\qq \color{green!20!blue} + 2 G_0(x-z) \pa^\m C(z-x) \, \pa \cdot \! A^a(y) n^\n \pa_{ \{\n} C(y-x) A_{\m \}}^a(x) \\
&\qq \color{green!20!blue} - 2 G_0(x-z) \pa^\m C(z-y) A_\n^a(y) n^\l \pa^\n \pa_{\{\l} C(y-x) A_{\m\}}^a(x) \\
&\qq \color{green!20!blue} - \frac{1-D}{2} \d(0) C(z-y) \, \pa \cdot \! A^a(y) C(z-x) \, \pa \cdot \! A^a(x)  \color{black} \bigg\} \, .
\end{aligned}
\end{align}
The blue terms cancel for any dimension $D$. Notice that this applies also to the formally divergent terms
including a factor of $\d(0)$ (which can be appropriately regularized).
The remaining black terms in \eqref{eq:Jac1} and \eqref{eq:Jac2} need to match the 
three terms from \eqref{eq:FPMSSDet}. One of them is identically satisfied,
while the two others are
\begin{align}
\begin{aligned}
\frac{r}{2} &= \frac{D}{2} - \frac{4-D}{2} = D-2 \, , \\
- \frac{r}{4} &= \frac{2 - D}{2} 
\end{aligned}
\end{align}
and are thus satisfied with $r=2(D-2)$. In particular this is true even without restricting to the gauge 
surface $n^\m A_\m^a = 0$. 
Let us further remark that we have also computed the Nicolai map in Landau gauge from the the rescaled field formalism. When performing the tests in Landau gauge we found that the determinants match either on the gauge surface $\pa^\m A_\m^a = 0$ for any $r=2(D-2)$ or everywhere else for $D=4$ only.

\section{Outlook}
In this paper we have presented explicit results for $\cT_g$ beyond the ones known
so far, and for different gauge choices. The fact that these are rather complicated is due 
to the fact that we have been considering {\em gauge-variant} expressions. We anticipate 
that the pertinent expressions will simplify substantially for the gauge-invariant operators 
that are usually considered in studies of $\cN=4$ Yang-Mills theory, as well as for 
the BPS-protected objects annihilated by the action of the $\cR$-operator. These topics
will be left for future study.

\vspace{0.2cm}
\noindent{\bf Acknowledgments:} We are grateful to the referee for helpful comments 
and for insisting on further clarification concerning the admissible gauge functionals.

\appendix
\section{Construction of the \texorpdfstring{$\tilde{\cR}$}{R}-operator}\label{sec:AppendixConstruction}
For the reader's convenience we here recall the derivation of the $\ctR$-operator in the 
rescaled field formalism \cite{Dietz:1984hf,Lechtenfeld:1984me},
pointing out the differences to the derivation of the $\cR$-operator in \cite{Ananth:2020lup}. In particular, we will see that unlike $\cR$, the $\ctR$-operator 
does not come with a multiplicative term $\langle\!\!\langle Z X \rangle\!\!\rangle$ 
which vanishes only on the gauge surface, and thus violates distributivity away from
this surface. Hence we will see that the $\ctR$-operator exists in any gauge. 
The full action $\tilde\rS = \tilde\rS_\text{inv} + \tilde\rS_\text{gf}$ is invariant under the 
BRST variations \eqref{eq:SlavnovVariations0} for all positive $\xi$ and arbitrary gauge-fixing 
functionals $\cG^a[\tA]$ (which for simplicity we assume not to depend on $g$). 

\noindent
As in \cite{Ananth:2020lup} we start from the flow equation
\begin{align}\label{eq:dXdg}
\frac{\md}{\md g} \left\langle \tX \right\rangle_g = 
\frac{\md}{\md g} \left\langle \!\! \!\left\langle \tX \right\rangle \!\!\! \right\rangle_g = 
\Big\langle \!\!\! \Big\langle \frac{\mathrm{d}\tX}{\md g} \Big\rangle \!\!\! \Big\rangle_g -
i \Big\langle\! \!\! \Big\langle \frac{\md(\tS_\text{inv} + \tS_\text{gf})}{\md g} \ \tX 
\Big\rangle\!\!\! \Big\rangle_g \eqqcolon \big\langle \ctR \, \tX \big\rangle_g \, .
\end{align}
Because the $g$ dependence appears only as an overall factor in $\tilde\rS =
\tilde\rS_\text{inv} + \tilde\rS_\text{gf}$ we have
\begin{align}
\frac{\md\tS_\text{inv}}{\md g} &= - \frac{2 \tS_\text{inv}}{g} = - \frac{2}{g^3} \, \d_\a \tilde{\D}_\a \, ,
\end{align}
where $\tilde\D_\a$ is defined in \eqref{eq:Delta} (with tildes); note that, being fermionic,
$\d_\a$ and $\D_\a$ {\em anti}-commute. By contrast, in
\cite{Ananth:2020lup} we needed an extra term on the r.h.s, which is not of the 
form of a supervariation, but which is absent here thanks to the auxiliary field. Thus 
\eqref{eq:dXdg} becomes
\begin{align}
\frac{\md}{\md g} \langle \tX \rangle_g =
\Big\langle \!\!\! \Big\langle \frac{\md \tX}{\md g} \Big\rangle \!\!\! \Big\rangle_g \,+\,
\frac{2i}{g^3} \big\langle \!\! \big\langle (\d_\a \tilde\D_\a) \tX \big\rangle \!\!\big\rangle_g \,+\, 
\frac{2i}{g} \big\langle \!\! \big\langle \tilde\rS_\text{gf} \tX \big\rangle \!\! \big\rangle_g \, .
\end{align}
We then continue as before and rewrite 
\begin{align}
 \big\langle \!\! \big\langle (\d_\a \tilde\D_\a) \tX \big\rangle \!\!\big\rangle_g =
 \big\langle \!\! \big\langle \d_\a \big(\tilde\D_\a \tX \big) \big\rangle \!\!\big\rangle_g \,+ \,
 \big\langle \!\! \big\langle \tilde\D_\a \d_\a\tX \big\rangle \!\!\big\rangle_g \, .
\end{align}
Next we use the supersymmetry Ward identity
\begin{align}
 \big\langle \!\! \big\langle \d_\a \tilde Y \big\rangle \!\!\big\rangle_g = -i 
 \big\langle \!\! \big\langle (\d_\a \tilde\rS_\text{gf}) \tilde Y \big\rangle \!\!\big\rangle_g \, .
\end{align}
Employing the Slavnov variations \eqref{eq:SlavnovVariations0} one finds that
\begin{align}
\tilde\rS_\text{gf} = - s \left( \frac{1}{g^2} \int \md x \ \bar{\tC}^a \cG^a[\tA]) \right) \, , 
\end{align}
which in particular implies
\begin{align}
\d_\a \tilde\rS_\text{gf} = - s \left(\frac{1}{g^2} \int \md x \ \bar{\tC}^a \d_\a \cG^a[\tA] \right) \, .
\end{align}
Thus, the Ward identity becomes
\begin{align}
\left< \!\! \left< \d_\a \, Y \right> \!\! \right>_g = -\, 
\Big\langle \!\!\! \Big\langle \frac{i}{g^2} \int \md x \ \bar{\tC}^a(x) \d_\a \cG^a(\tA) \ 
s(\tilde Y) \Big\rangle \!\!\! \Big\rangle_g \, .
\end{align}
We now apply this identity to $\tilde Y = \tilde{\D}_\a \tX$. Because $\tilde{\D}_\a$ is gauge 
invariant we have $s(\tilde{\D}_\a) = 0$ and thus $s(\tilde{\D}_\a X) = - \tilde{\D}_\a s(\tX)$
(the minus sign here appears because $s$ anti-commutes with fermionic expressions).
Subsequently we put everything back together to obtain
\begin{align}
\begin{gathered}\label{eq:Rop1}
\frac{\md}{\md g} \langle\!\!\langle \tX \rangle\!\!\rangle_g
 = \Big\langle \!\!\!\Big\langle \frac{\md \tX}{\md g} \Big\rangle \!\!\!\Big\rangle_g 
 \, -\, \frac{2i}{g^3} \big\langle \!\! \big\langle \tilde{\D}_\a \d_\a \, \tX \big\rangle \!\!\big\rangle_g 
 \,-\, \frac{2}{g^5} \Big\langle \!\! \!\Big\langle \int \md x \ \bar{\tC}^a(x) \d_\a \cG^a(\tA) \, 
 \tilde{\D}_\a \, s(\tX) \Big\rangle \!\!\!\Big\rangle_g \\
+ \frac{2}{g^3} \Big\langle \!\! \!\Big\langle \int \md x \ \bar{\tC}^a(x) \cG^a(\tA) 
\, s(\tX) \Big\rangle \!\!\!\Big\rangle_g \, .
\end{gathered}
\end{align}
Unlike the $\cR$-operator constructed in \cite{Ananth:2020lup} the r.h.s. of \eqref{eq:Rop1} 
does not contain a multiplicative contribution which only vanishes on the gauge surface,
and therefore acts distributively without further ado, and for any $\cG^a[\tA]$.
Finally we integrate \eqref{eq:Rop1} over all fermionic degrees of freedom. Each integration 
absorbs two powers of $\textstyle\frac{1}{g}$, so we arrive at
\begin{align}
\begin{aligned}\label{eq:Rop2}
\ctR \, \tX = \frac{\mathrm{d}X}{\md g} \,+\,\frac{2i}{g} 
\bcontraction{}{\d}{_\a \tX \ }{\tilde{\D}}
\d_\a \tX \ \tilde{\D}_\a - 
\frac{2}{g} \int \md x \ 
\bcontraction[2ex]{}{\bar{\tC}}{^a(x) \d_\a \cG^a(\tA) \tilde{\D}_\a }{s}
\bar{\tC}^a(x) 
\bcontraction{}{\d}{_\a \cG^a(\tA) }{\tilde{\D}}
\d_\a \cG^a(\tA) \tilde{\D}_\a s(\tX) + \frac{2}{g} \int \md x \ 
\bcontraction{}{\bar{\tC}}{^a(x) \cG^a(\tA) }{s}
\bar{\tC}^a(x) \cG^a(\tA) s(\tX) 
\end{aligned}
\end{align}
from which the distributivity of $\ctR$ is manifest by the distributivity of
$\d_\a$ and $s$. The final form \eqref{eq:RFinal} is arrived at by taking
$\tX = \tA_\mu^a$ with $s(\tA_\mu^a) = (\tilde{\mD}_\mu \tC)^a$ and substituting 
the formulas for the ghost and gaugino propagators.

\noindent
For gauge invariant $\tX$ the above formula reduces to
\begin{align}
\ctR \tX \,\equiv\, 
\ctR_\text{inv} \tX \,:=\, \frac{\mathrm{d}\tX}{\md g} +
\frac{2i}{g} \bcontraction{}{\d}{_\a \tX \ }{\tilde{\D}}
\d_\a \tX \ \tilde{\D}_\a \, .
\end{align}
A straightforward calculation analogous to the one in A.3 of \cite{Ananth:2020lup} shows that 
\begin{align}
-\frac{1}{4g^2} \, \int \md x \ \tF_{\m\nu}^a \tF^{a\, \m\nu} \q \text{and} 
\q -\frac{1}{2g^2} \, \int \md x \ \tD^a \tD^a 
\end{align}
are in the kernel of $\ctR$. Thus we can set $\tD^a = 0$ without loss of generality. 
Using the definitions of the fermion and ghost propagators
\begin{align}
i \bcontraction{}{\tl}{^a(x) }{\btl} \tl^a(x) \btl^b(y) = \tS^{ab}(x,y;\tA) \q \text{and} \q \bcontraction{}{\tC}{^a(x) }{\bar{\tC}} \tC^a(x) \bar{\tC}^b(y) = \tG^{ab}(x,y;\tA) 
\end{align}
we obtain the final form of $\ctR = \ctR_\text{inv} + \ctR_\text{gf}$ 
spelled out in \eqref{eq:RInvFinalR} and \eqref{eq:RGfFinalR}.

\section{Fourth order result in Landau gauge}
In this appendix we give the explicit form of the on-shell map $\cT_g$ for the Landau gauge up 
to and including order $\cO(g^4)$, thus extending the result of \cite{Ananth:2020lup}
by one order. This expression does satisfy all the tests on-shell, that is, on the
gauge surface $\pa^\mu A_\mu^a = 0$; details of the latter calculation
will be provided in a forthcoming thesis \cite{HM}. We  emphasize that for the $\cN=1$
theory our prescription would also yield the corresponding off-shell result, but
the resulting expressions would be considerably more cumbersome.

\begin{footnotesize}
\begin{align}
\begin{aligned} 
(\cT_gA)_\m^a(x) &= A_\m^a(x) + g\,f^{abc} \int \md y \ \pa^\r C(x-y) A_\m^b(y) A_\r^c(y) \\
&\q + \, \frac{3g^2}{2} f^{abc} f^{bde} \int \md y \ \md z \ \pa^\r C(x-y) A^{\l \, c}(y) \pa_{[\r} C(y-z) A_\m^d(z) A_{\l]}^e(z) \\
&\q + \frac{g^3}{2} f^{a b c} f^{b d e} f^{c f g} \int \md y \ \md z \ \md w \ \pa^\r C(x-y) \\
&\qqq \times \pa^\l C(y-z) A_\l^d(z) A^{\s \, e}(z) \pa_{[\r} C(y-w) A_\m^f(w) A_{\s]}^g(w) \\
&\q + \, g^3 f^{a b c} f^{b d e} f^{d f g} \int \md y \ \md z \ \md w \ \pa^\r C(x-y) A^{\l \, c}(y) \ \bigg\{ \\
&\qqq - \pa^{\s} C(y-z)A_{\s}^e(z) \pa_{[\r} C(z-w) A_\m^f(w) A_{\l]}^g(w) \\
&\qqq + \pa_{[\r} C(y-z) A_\m^e(z) \pa^\s C(z-w) A_{\l ]}^f(w) A_\s^g(w) \bigg\} \\
&\q + \, \frac{g^3}{3} f^{abc} f^{bde} f^{dfg} \int \md y \ \md z \ \md w \ \bigg\{ \\
&\qqq + 6 \, \pa_\r C(x-y) A^{\l \, c}(y) \pa^{[\r} C(y-z) A^{\s] \, }(z) \pa_{[\l} C(z-w) A_\m^f(w) A_{\s]}^g(w) \\
&\qqq - 6 \, \pa^\r C(x-y) A_\l^c(y) \pa^{[\l} C(y-z) A^{\s] \, e}(z) \pa_{[\r} C(z-w) A_\m^f(w) A_{\s]}^g(w) \\
&\qqq - 6 \, \pa_\r C(x-y) A_\l^c(y) \pa_{[\s} C(y-z) A_{\m]}^e(z) \pa^{[\r} C(z-w) A^{\l \, f}(w) A^{\s] \, g}(w) \\
&\qqq + 2 \, \pa^\r C(x-y) A_{[\r}^c(y) \pa_{\m]} C(y-z) A^{\l \, e}(z) \pa^\s C(z-w) A_\l^f(w) A_\s^g(w) \\
&\qqq - \pa_\m C(x-y) \, \pa^\r \left( A_\r^c(y) C(y-z) \right) A^{\l \, e}(z) \pa^\s C(z-w) A_\l^f(w) A_\s^g(w) \bigg\} \\
&\q - \, \frac{g^3}{3} f^{abc} f^{bde} f^{dfg} \int \md y \ \md z 
 \ A_\m^c(x) C(x-y) A^{\r \, e}(y) \pa^\l C(y-z) A_\r^f(z) A_\l^g(z) \\
 &\q + \frac{g^4}{12} f^{a b c} f^{b d e} f^{d f g} f^{c h i} \int \md y \ \md z \ \md w \\
&\qqq \times C(x-y) A^{\l \, e}(y) \pa^\r C(y-z) A_\l^f(z) A_\r^g(z) \pa^\s C(x-w) A_\s^h(w) A_\m^i(w) \\
&\q + \frac{g^4}{8} f^{a b c} f^{b d e} f^{d f g} f^{c h i} \int \md y \ \md z \ \md w \ \md v \ \pa^\l C(x-y) \pa_\r C(y-z) \ \bigg\{ \\
&\qqq - 9 A_\s^e(z) \pa^{[\r} C(z-w) A^{\s \, f}(w) A^{\n] \, g}(w) \pa_{[\m} C(y-v) A_\l^h(v) A_{\n]}^i(v) \\ 
&\qqq + 4 A^{[\r \, e}(z) \pa^{|\s|} C(z-w) A_\s^f(w) A^{\n] \, g}(w) \pa_{[\m} C(y-v) A_\l^h(v) A_{\n]}^i(v) \\
&\qqq - 2 A^{\r \, e}(z) \pa_{[\m} C(z-w) A_\l^f(w) A_{\n]}^g(w) \pa^\s C(y-v) A_\s^h(v) A^{\n \, i}(v) \bigg\} 
\end{aligned}
\end{align}
\newpage
\begin{align*}
\hphantom{(\cT_gA)_\m^a(x)} &\q - \frac{g^4}{12} f^{a b c} f^{b d e} f^{d f g} f^{c h i} \int \md y \ \md z \ \md w \ \md v \ \pa_\m C(x-y) \pa^\l C(y-z) \\
&\qqq \times A^{\r \, e}(z) \pa^\s C(z-w) A_\s^f(w) A_\r^g(w) \pa^\tau C(y-v) A_\tau^h(v) A_\l^i(v) \\ 
&\q + \frac{g^4}{2} f^{a b c} f^{b d e} f^{d f g} f^{c h i} \int \md y \ \md z \ \md w \ \md v \ \pa_\l C(x-y) \ \bigg\{ \\
&\qqq + \pa_{[\m} C(y-z) A_{\r]}^e(z) \pa^{[\l} C(z-w) A^{\r \, f}(w) A^{\n] \, g}(w) \pa^\s C(y-v) A_\s^h(v) A_\n^i(v) \\ 
&\qqq - \pa^{[\l} C(y-z) A^{\r] \, e}(z) \pa_{[\m} C(z-w) A_\r^f(w) A_{\n]}^g(w) \pa^\s C(y-v) A_\s^h(v) A_\n^i(v) \bigg\} \\
&\q + \frac{g^4}{6} f^{a b c} f^{b d e} f^{d f g} f^{c h i} \int \md y \ \md z \ \md w \ \md v \ \pa^\l C(x-y) \ \bigg\{ \\
&\qqq + 3 \pa^{[\r} C(y-z) A^{\n] \, e}(z) \pa_{[\m} C(z-w) A_\l^f(w) A_{\r]}^g(w) \pa^\s C(y-v) A_\s^h(v) A_\n^i(v) \\ 
&\qqq + \pa_{[\l} C(y-z) A^{\n \, e}(z) \pa^\s C(z-w) A_{|\s}^f(w) A_\n^g(w) \pa^\r C(y-v) A_{\r|}^h(v) A_{\m]}^i(v) \bigg\} \\
&\q - \frac{g^4}{3} f^{a b c} f^{b d e} f^{d f g} f^{e h i} \int \md x \ \md y \ \md z \ \md w \\
&\qqq \times A_\m^c(x) C(x-y) \pa^\l C(y-z) A_\l^f(z) A^{\r \, g}(z) \pa^\s C(y-w) A_\s^h(w) A_\r^i(w) \\
&\q - \frac{g^4}{3} f^{a b c} f^{b d e} f^{d f g} f^{e h i} \int \md y \ \md z \ \md w \ \md v \ \pa_\m C(x-y) \\
&\qqq \times \pa^\l \left( A_\l^c(y) C(y-z) \right) \pa^\r C(z-w) A_\r^f(w) A^{\n\, g}(w) \pa^\s C(z-v) A_\s^h(v) A_\n^i(v) \\
&\q + \frac{g^4}{12} f^{a b c} f^{b d e} f^{d f g} f^{e h i} \int \md y \ \md z \ \md w \ \md v \ \pa^\l C(x-y) A^{\r \, c}(y) \ \bigg\{ \\
&\qqq - 3 \pa_\r C(y-z) \pa_{[\m} C(z-w) A_\l^f(w) A_{\n]}^g(w) \pa^\s C(z-v) A_\s^h(v) A_\n^i(v) \\ 
&\qqq - 3 \pa^\n C(y-z) \pa_{[\m} C(z-w) A_\r^f(w) A_{\n]}^g(w) \pa^\s C(z-v) A_\s^h(v) A_\l^i(v) \\ 
&\qqq + 3 \pa^\n C(y-z) \pa_{[\l} C(z-w) A_\r^f(w) A_{\n]}^g(w) \pa^{\s} C(z-v) A_{\s}^h(v) A_\m^i(v) \\ 
&\qqq - 3 \pa_{\m} C(y-z) \pa_{[\l} C(z-w) A_\r^f(w) A_{\n]}^g(w) \pa^{\s} C(z-v) A_{\s}^h(v) A_\n^i(v) \\ 
&\qqq + 3 \pa_{\l} C(y-z) \pa_{[\m} C(z-w) A_\r^f(w) A_{\n]}^g(w) \pa^{\s} C(z-v) A_{\s}^h(v) A^{\n \, i}(v) \\ 
&\qqq - 2 \pa_{[\l} C(y-z) \pa^\n C(z-w) A_{|\n|}^f(w) A_{\m]}^g(w) \pa^{\s} C(z-v) A_\s^h(v) A_\r^i(v) \\ 
&\qqq + \pa_{\r} C(y-z) \pa^\n C(z-w) A_{\n}^f(w) A_{\m}^g(w) \pa^\s C(z-v) A_\s^h(v) A_\l^i(v) \bigg\} \\ 
&\q + \frac{g^4}{6} f^{a b c} f^{b d e} f^{d f g} f^{e h i} \int \md y \ \md z \ \md w \ \md v \ \pa^\l C(x-y) \ \bigg\{ \\
&\qqq - 7 A_{[\m}^c(y) \pa_{\l]} C(y-z) \pa^\r C(z-w) A_\r^f(w) A^{\n \, g}(w) \pa^\s C(z-v) A_\s^h(v) A_\n^i(v) \\ 
&\qqq + 3 A^{[\n \, c}(y) \pa^{\r]} C(y-z) \pa_{[\m} C(z-w) A_\l^f(w) A_{\r]}^g(w) \pa^\s C(z-v) A_\s^h(v) A_\n^i(v) \bigg\} \\
&\q - \frac{g^4}{2} f^{a b c} f^{b d e} f^{d f g} f^{f h i} \int \md y \ \md z \ \md w \ \\
&\qqq \times \pa^\l C(x-y) A_{[\m}^c(y) A_{\l]}^e(y) C(y-z) A^{\r \, g}(z) \pa^\s C(z-w) A_\s^h(w) A_\r^i(w) 
\end{align*}
\newpage
\begin{align*}
\hphantom{(\cT_gA)_\m^a(x)} &\q + \frac{g^4}{12} f^{a b c} f^{b d e} f^{d f g} f^{f h i} \int \md x \ \md y \ \md z \ \md w \ A_\m^c(x) C(x-y) \ \bigg\{ \\
&\qqq + 9 A^{\l \, e}(y) \pa^\r C(y-z) A^{\s \, g}(z) \pa_{[\l} C(z-w) A_\r^h(w) A_{\s]}^i(w) \\ 
&\qqq + 4 A^{[\l \, e}(y) \pa^{\r]} C(y-z) A_\l^g(z) \pa^\s C(z-w) A_\s^h(w) A_\r^i(w) \\
&\qqq - 3 A_\l^e(y) \pa^\l C(y-z) A^{\r \, g}(z) \pa^\s C(z-w) A_\s^h(w) A_\r^i(w) \\ 
&\qqq - 3 \pa^\l \left( A_\l^e(y) C(y-z) \right) A^{\r \, g}(z) \pa^\s C(z-w) A_\s^h(w) A_\r^i(w) \bigg\} \\ 
&\q + \frac{g^4}{12} f^{a b c} f^{b d e} f^{d f g} f^{f h i} \int \md y \ \md z \ \md w \ \md v \ \pa_\m C(x-y) \pa^\l \left( A_\l^c(y) C(y-z) \right) \ \bigg\{ \\
&\qqq + 9 A^{\r \, e}(z) \pa^\s C(z-w) A^{\n \, g}(w) \pa_{[\r} C(w-v) A_\s^h(v) A_{\n]}^i(v) \\ 
&\qqq + 4 A^{[\r \, e}(z) \pa^{\n]} C(z-w) A_\r^g(w) \pa^\s C(w-v) A_\s^h(v) A_\n^i(v) \\
&\qqq - 3 \pa^\r \left( A_\r^e(z) C(z-w) \right) A^{\n \, g}(w) \pa^\s C(w-v) A_\s^h(v) A_\n^i(v) \\ 
&\qqq - 3 A_\r^e(z) \pa^\r C(z-w) A^{\n \, g}(w) \pa^\s C(w-v) A_\s^h(v) A_\n^i(v) \bigg\} \\ 
&\q + \frac{g^4}{2} f^{a b c} f^{b d e} f^{d f g} f^{f h i} \int \md y \ \md z \ \md w \ \md v \ \pa^\l C(x-y) A_{[\m}^c(y) \pa_{\l]} C(y-z) \ \bigg\{ \\
&\qqq - \pa^\r \left( A_\r^e(z) C(z-w) \right) A^{\n \, g}(w) \pa^\s C(w-v) A_\s^h(v) A_\n^i(v) \\ 
&\qqq - A_\r^e(z) \pa^\r C(z-w) A^{\n \, g}(w) \pa^\s C(w-v) A_\s^h(v) A_\n^i(v) \bigg\} \\
&\q + \frac{2g^4}{3} f^{a b c} f^{b d e} f^{d f g} f^{f h i} \int \md y \ \md z \ \md w \ \md v \ \pa^\l C(x-y) \\
&\qqq \times A_{[\m}^c(y) \pa_{\l]} C(y-z) A^{\r \, e}(z) \pa^\n C(z-w) A_{[\r}^g(w) \pa^\s C(w-v) A_{|\s|}^h(v) A_{\n]}^i(v) \\
&\q + \frac{3g^4}{2} f^{a b c} f^{b d e} f^{d f g} f^{f h i} \int \md y \ \md z \ \md w \ \md v \ \pa_\l C(x-y)  \ \bigg\{ \\
&\qqq + 4 A^{\r\, c}(y) \pa^{[\l} C(y-z) A^{e \, \n]}(z) \pa^\s C(z-w) A_{[\m}^g(w) \pa_\r C(w-v) A_\s^h(v) A_{\n]}^i(v) \\
&\qqq - 4 A_r^c(y) \pa_{[\m} C(y-z) A_{\n]}^e(z) \pa_\s C(z-w) A^{g\, [\l}(w) \pa^\r C(w-v) A^{h \, \s}(v) A^{i \, \n]}(v) \\
&\qqq - A^{\r\, c}(y) \pa_{[\r} C(y-z) A_{\s]}^e(z) \pa_\m C(z-w) A_\n^g(w) \pa^{[\l} C(w-v) A^{h \, \s}(v) A^{i \, \n]}(v) \bigg\} \\
&\q + \frac{3g^4}{2} f^{a b c} f^{b d e} f^{d f g} f^{f h i} \int \md y \ \md z \ \md w \ \md v \ \pa^\l C(x-y) \ \bigg\{ \\
&\qqq - A_{[\l}^c(y) \pa_{\m]} C(y-z) A^{\r \, e}(z) \pa^\s C(z-w) A^{\n \, g}(w) \pa_{[\r} C(w-v) A_\s^h(v) A_{\n]}^i(v) \\
&\qqq - A^{[\r \, c}(y) \pa_\l C(y-z) A^{\s] \, e}(z) \pa_\r C(z-w) A^{\n \, g}(w) \pa_{[\m} C(w-v) A_\s^h(v) A_{\n]}^i(v) \\
&\qqq + A^{[\r \, c}(y) \pa_\m C(y-z) A^{\s] \, e}(z) \pa_\r C(z-w) A^{\n \, g}(w) \pa_{[\l} C(w-v) A_\s^h(v) A_{\n]}^i(v) \\
&\qqq + A^{[\r \, c}(y) \pa^{\s]} C(y-z) A_\l^e(z) \pa_\r C(z-w) A^{\n \, g}(w) \pa_{[\m} C(w-v) A_\s^h(v) A_{\n]}^i(v) \\
&\qqq - A^{[\r \, c}(y) \pa^{\s]} C(y-z) A_\m^e(z) \pa_\r C(z-w) A^{\n \, g}(w) \pa_{[\l} C(w-v) A_\s^h(v) A_{\n]}^i(v) \bigg\} 
\end{align*}
\newpage
\begin{align*}
\hphantom{(\cT_gA)_\m^a(x)} &\q + \frac{g^4}{2} f^{a b c} f^{b d e} f^{d f g} f^{f h i} \int \md y \ \md z \ \md w \ \md v \ \pa^\l C(x-y) A^{\r \, c}(y) \ \bigg\{ \\
&\qqq - 8 \pa^\n C(y-z) A_{[\m}^e(z) \pa_\l C(z-w) A_\r^g(w) \pa^\s C(w-v) A_{|\s|}^h(v) A_{\n]}^i(v) \\
&\qqq - 2 \pa_\l C(y-z) A^{\n \, e}(z) \pa_{[\m} C(z-w) A_\r^g(w) \pa^\s C(w-v) A_{|\s|}^h(v) A_{\n]}^i(v) \\ 
&\qqq - 2 \pa_\r C(y-z) A^{\n \, e}(z) \pa_{[\m} C(z-w) A_\l^g(w) \pa^\s C(w-v) A_{|\s|}^h(v) A_{\n]}^i(v) \\ 
&\qqq + 2 \pa_\m C(y-z) A^{\n \, e}(z) \pa_{[\l} C(z-w) A_\r^g(w) \pa^\s C(w-v) A_{|\s|}^h(v) A_{\n]}^i(v) \\ 
&\qqq - \frac{3}{2} \pa_{[\m} C(y-z) A_{\r]}^e(z) \pa^\s C(z-w) A^{\n \, g}(w) \pa_{[\l} C(w-v) A_\s^h(v) A_{\n]}^i(v) \\ 
&\qqq + \frac{3}{2} \pa_{[\l} C(y-z) A_{\r]}^e(z) \pa^\s C(z-w) A^{\n \, g}(w) \pa_{[\m} C(w-v) A_\s^h(v) A_{\n]}^i(v) \\ 
&\qqq + \frac{3}{2} \pa_{[\m} C(y-z) A_{\l]}^e(z) \pa^\s C(z-w) A^{\n \, g}(w) \pa_{[\r} C(w-v) A_\s^h(v) A_{\n]}^i(v) \\ 
&\qqq + \pa_{[\m} C(y-z) A_\l^e(z)  \pa^\n C(z-w) A_{\r]}^g(w) \pa^\s C(w-v) A_\s^h(v) A_\n^i(v) \\ 
&\qqq - \pa_{[\m} C(y-z) A_\l^e(z) \pa^\n C(z-w) A_{|\n}^g(w) \pa^\s C(w-v) A_{\s|}^h(v) A_{\r]}^i(v) \bigg\} \\ 
&\q + \frac{3g^4}{4} f^{a b c} f^{b d e} f^{d f g} f^{f h i} \int \md y \ \md z \ \md w \ \md v \ \pa^\l C(x-y) A^{\r\, c}(y) \ \bigg\{ \\
&\qqq - 20 \pa^\n C(y-z) A^{\s \, e}(z) \pa_{[\n} C(z-w) A_\m^g(w) \pa_\s C(w-v) A_\l^h(v) A_{\r]}^i(v) \\
&\qqq - 4 \pa_\r C(y-z) A^{\n \, e}(z) \pa^\s C(z-w) A_{[\m}^g(w) \pa_\l C(w-v) A_\s^h(v) A_{\n]}^i(v) \\
&\qqq + 4 \pa^\s C(y-z) A_\s^e(z) \pa^\n C(z-w) A_{[\m}^g(w) \pa_\l C(w-v) A_\r^h(v) A_{\n]}^i(v) \\
&\qqq - 2 \pa_{[\m} C(y-z) A^{\s \, e}(z) \pa_{\l]} C(z-w) A^{\n \, g}(w) \pa_{[\r} C(w-v) A_\s^h(v) A_{\n]}^i(v) \\
&\qqq + 2 \pa^\s C(y-z) A_{[\m}^e(z) \pa_{\l]} C(z-w) A^{\n \, g}(w) \pa_{[\r} C(w-v) A_\s^h(v) A_{\n]}^i(v) \\
&\qqq - 2 \pa^\s C(y-z) A_{[\s}^e(z) \pa_{\r]} C(z-w) A^{\n \, g}(w) \pa_{[\m} C(w-v) A_\l^h(v) A_{\n]}^i(v) \\
&\qqq - 2 \pa_{[\m} C(y-z) A_\l^e(z) \pa_{\r]} C(z-w) A^{\n \, g}(w) \pa^\s C(w-v) A_\s^h(v) A_\n^i(v) \\ 
&\qqq - 2 \pa^\s C(y-z) A_\r^e(z) \pa^\n C(z-w) A_{[\m}^g(w) \pa_\l C(w-v) A_{\s]}^h(v) A_\n^i(v) \\
&\qqq - \pa_\r C(y-z) A_\s^e(z) \pa^\s C(z-w) A^{\n \, g}(w) \pa_{[\m} C(w-v) A_\l^h(v) A_{\n]}^i(v) \\
&\qqq - \pa^\s C(y-z) A_\s^e(z) \pa_\m C(z-w) A^{\n \, g}(w) \pa_{[\l} C(w-v) A_\r^h(v) A_{\n]}^i(v) \\
&\qqq - \pa^\s C(y-z) A_\r^e(z) \pa^\n C(z-w) A_{[\m}^g(w) \pa_{|\n|} C(w-v) A_\l^h(v) A_{\s]}^i(v) \\ 
&\qqq + \pa^\s C(y-z) A_\r^e(z) \pa^\n C(z-w) A_\n^g(w) \pa_{[\m} C(w-v) A_\l^h(v) A_{\s]}^i(v) \\
&\qqq + \pa^\s C(y-z) A_\s^e(z) \pa_\l C(z-w) A^{\n \, g}(w) \pa_{[\m} C(w-v) A_\r^h(v) A_{\n]}^i(v) \bigg\} \\
&\q + \cO(g^5) \, .
\end{align*}
\end{footnotesize}

\end{document}